\documentclass[preprint,eqsecnum,aps,showpacs,floatfix]{revtex4}

\usepackage{graphicx}

\usepackage{amsmath,amssymb}

\usepackage{graphicx,epsfig}

\usepackage{dcolumn}
\usepackage{bm}

\newcommand{\Z}{{\mathbb Z}}

\begin{document}


\title{Gauged Baryon and Lepton Number in MSSM$_4$ Brane Worlds}

\author{Richard F. Lebed}
\email{Richard.Lebed@asu.edu}

\author{Van E. Mayes}
\email{Van.Mayes@asu.edu}

\affiliation{Department of Physics, Arizona State University,
Tempe, AZ 85287-1504}

\date{June, 2011}

\begin{abstract}
A recent D-brane model designed to accommodate a phenomenologically
acceptable fourth generation of chiral fermions was noted to produce
an unexpected additional unbroken nonanomalous U(1) gauge group at the
string scale.  We show that the corresponding charges acting on MSSM
fields count baryon and lepton numbers.  If broken spontaneously at
lower scales, these U(1)$_B$ and U(1)$_L$ symmetries provide potential
avenues for preserving baryogenesis while nonetheless explaining the
suppression of proton decay (without the need for $R$ parity), as well
as the smallness of right-handed neutrino Majorana masses compared to
the string scale.
\end{abstract}


\maketitle
\thispagestyle{empty}

\newpage
\setcounter{page}{1}

\section{Introduction} \label{sec:Intro}

The long lifetime of the proton, in excess of $10^{34}$~y, provides
one of the strongest hints of fundamental physics beyond the
electroweak scale.  Baryon ($B$) and lepton ($L$) number are conserved
as global symmetries of the Standard Model (SM) and the Minimal
Supersymmetric Standard Model (MSSM).  However, global symmetries in a
theory can easily appear by accident, and unless they are protected in
some way, are broken at some level by nonperturbative effects.
Indeed, one of the distinctive features of grand unified theories
(GUTs) is their introduction of new particles that can mediate proton
decay.  A typical non-supersymmetric GUT with arbitrary couplings of
natural size allows proton decay at an unacceptably high rate, which
[in addition to the taming of the gauge hierarchy problem and the
numerical improvement of SU(3)$_C \times$SU(2)$_L
\times$U(1)$_Y$ gauge unification] provides impetus for the study of
supersymmetric (SUSY) theories.  In particular, the scale at which 
unification occurs in SUSY models is pushed to a slightly higher 
energy, leading to an additional suppression of the proton decay rate. 

However, unmodified SUSY theories generate their own problems with $B$
and $L$ nonconservation.  While $B$- and $L$-violating terms in
non-SUSY theories first appear in operators of dimension $d = 6$, the
presence of squark and slepton fields allows such terms to arise
already at $d = 4$.  Denoting the left-handed chiral matter and Higgs
superfields by the quantum numbers of $B$, $L$, and SU(3)$_C
\times$SU(2)$_L \times$U(1)$_Y$ in the usual way,
\begin{eqnarray} \label{chiral} & &
Q ({\scriptstyle +\frac 1 3}, 0, {\bf 3}, {\bf 2},
{\scriptstyle +\frac 1 6}) \, , \
L (0, +1, {\bf 1}, {\bf 2}, {\scriptstyle -\frac 1 2}) \, , \
U^c ( {\scriptstyle -\frac 1 3}, 0, \overline{\bf 3}, {\bf 1},
{\scriptstyle -\frac 2 3}) \, , \
D^c ( {\scriptstyle -\frac 1 3}, -1, \overline{\bf 3}, {\bf 1},
{\scriptstyle +\frac 1 3}) \, , \nonumber \\ & &
N^c ( -1, 0, {\bf 1}, {\bf 1}, 0) \, , \
E^c ( -1, 0, {\bf 1}, {\bf 1}, +1) \, , \
H_u (0, 0, {\bf 1}, {\bf 2}, {\scriptstyle +\frac 1 2}) \, , \
H_d (0, 0, {\bf 1}, {\bf 2}, {\scriptstyle -\frac 1 2}) \, , \
\end{eqnarray}
one finds already at $d = 4$ several $B$- and $L$-violating operators,
such as $N^c N^c N^c$ and $Q D^c L$ (where the necessary color and
weak isospin contractions are implied).  Conventional $R$
parity~\cite{Dimopoulos:1981zb,Sakai:1981gr,
Sakai:1981pk,Nilles:1981py,Dimopoulos:1981dw}, which may be defined
operationally on MSSM fields as $(-1)^{3(B-L)+2J}$, eliminates such
terms while protecting the conventional Yukawa terms such as $L H_u
N^c$.  Nevertheless, even with this restriction
Refs.~\cite{Sakai:1981pk,Weinberg:1981wj} showed that $d = 5$
operators such as $QQQL$ still lead to an unacceptably short proton
lifetime unless additional suppressions are introduced.

A classic study~\cite{Ibanez:1991pr} of possible $B$- and
$L$-violating operators at $d = 4$ and 5 and how various discrete
symmetries may be used to suppress phenomenologically unacceptable
(``dangerous'') decay rates also summarizes which of these operators
are especially dangerous by providing estimates of their suppressed
coefficients.  While the number of such operators increases greatly
after allowing the right-handed neutrino field $N^c$~\cite{Lee:2007fw}
(which was not considered in~\cite{Ibanez:1991pr}; we count 36 $B$-
and/or $L$-violating operators up to and including $d = 5$), one can
check that none of them are singlets under a symmetry that assigns a
charge $Q_X \! = \! -1$ to $Q$ and $L$, $+1$ to $U^c$, $D^c$, $N^c$,
and $E^c$, and 0 to $H_{u,d}$ [for example, such a symmetry can occur
as a subgroup of a hypothetical U(1)$_X$, as seen below].  On the
other hand, the $\mu$ term $H_u H_d$ and the standard Yukawa terms are
all singlets under this symmetry, indicating a potential route for
suppressing proton decay without recourse to $R$ parity.

Going beyond global symmetries, one may consider gauging the
symmetries U(1)$_B$ or U(1)$_L$ that count baryon and lepton numbers,
respectively, either for its own phenomenological interest in collider
physics~\cite{Carone:1994aa}, or as a method of providing interesting
dark matter candidates in both non-SUSY~\cite{FileviezPerez:2010gw}
and SUSY~\cite{Ko:2010at} models.  So-called leptophilic matter
[having a U(1) charge that couples preferentially to leptons rather
than quarks, but not necessarily just U(1)$_L$] can serve to
explain~\cite{Fox:2008kb} the observed
PAMELA~\cite{Adriani:2008zr}/ATIC~\cite{Chang:2008zzr}/Fermi
LAT~\cite{Ackermann:2011rq} cosmic ray positron excess, while a
relatively light leptophobic [coupling primarily to quarks rather than
leptons, but not necessarily just U(1)$_B$] $Z^\prime$ can serve to
explain~\cite{Buckley:2011vc} Tevatron anomalies in the measured $t
\bar t$ forward-backward asymmetry~\cite{Aaltonen:2011kc} and the
associated production of $W$s with jets~\cite{Aaltonen:2011mk}.
Introducing additional Abelian symmetries beyond those of the SM or
MSSM requires the consideration of a variety of new potential
anomalies and the addition of new chiral matter multiplets to enforce
their cancellation; addressing such concerns accounts for much of the
analysis in the models mentioned above.

The introduction of almost any extra gauged U(1) that contains a
portion coupling universally to quarks [{\it i.e.}, U(1)$_B$] leads to
the prohibition of proton decay at energy scales such that the
symmetry remains unbroken.  Since the proton is the lightest fermionic
hadron, its decay products must contain an odd number of leptons, and
some potential combinations including U(1)$_B$, such as U(1)$_{B-L}$
(if present), would have to break at a somewhat higher scale than
others in order to support the experimental lifetime limit.  Indeed,
such possibilities have been considered for
decades~\cite{Font:1989ai}.  However, demanding baryon number to be an
exactly conserved quantity at all scales would also eliminate the
possibility of baryogenesis and a natural resolution of the problem of
matter-antimatter asymmetry in the universe.

Analogous to the possibility that unconstrained operators can generate
excessively large proton decay but their complete elimination forbids
mechanisms for generating baryogenesis, the $\mu$-term coupling also
introduces a well-known difficulty~\cite{Dimopoulos:1981zb} in the
MSSM: The natural Planck- or string-scale value of $\mu$ would force
comparable values for the Higgs masses, and therefore must be
suppressed.  However, if $\mu$ is taken too small, light charginos
appear in mass ranges that have already been excluded by collider
searches.  In the original solution~\cite{Dimopoulos:1981zb}, SUSY is
explicitly but softly broken by the $\mu$ term in the TeV range;
however, one may also consider the possibility of $\mu$ arising from
spontaneous symmetry breaking.  An efficient model of beyond-MSSM
physics addresses both issues, as in
Refs.~\cite{Lee:2007fw,Aoki:1999tv}.  Indeed, both of these papers
also note (as suggested above) a diminished significance for $R$
parity.  Moreover, another prime motivation for $R$ parity---the
existence of a lightest SUSY particle as a dark matter candidate---is
ameliorated if one has recourse to other selection rules based upon
additional U(1) symmetries to produce other long-lived exotic
particles.

Under what circumstances can one develop theories with nonanomalous
U(1)s that help to stabilize the proton?  Here too, one can find
interesting observations in the literature.  As noted in
Refs.~\cite{Pati:1996fn,Barr:2005je}, the desired U(1)s do not appear
to fit into conventional SO(10) or $E_6$-type grand unification
schemes.  Amusingly, the former paper (by Pati) advocates a stringy
origin for the extra U(1), while the latter paper (inspired by results
of explicit models~\cite{Barr:1986hj,Ma:2001kg}) advocates a
Pati-Salam (left-right symmetric) scheme~\cite{Pati:1974yy} as the
origin of the extra U(1).  The model we present here, as it turns out,
supports both stringy and Pati-Salam ideas.  The possibility of
suppressing proton decay using a left-right symmetric heterotic
string-derived U(1) is considered in
Refs.~\cite{Coriano:2007ba,Faraggi:2011xu}, and a fourth generation in
heterotic string theory is considered in
Refs.~\cite{Khlopov:2002gg,Belotsky:2004ga}.

Furthermore, cancelling the anomalies of the U(1) associated with the
suppression of proton decay appears to be more natural in models that
introduce an additional set of chiral fermions;
Ref.~\cite{Aoki:1999tv} includes a full family of fermions with
opposite chiralities to those of the SM, but also notes that similar
cancellations are possible in a four-generation MSSM\@.  The potential
significance of four generations in this context is also explored in
Refs.~\cite{Perez:2011dg,Smith:2011rp}.  This is precisely the
scenario we propose here.  While long considered phenomenologically
unlikely, the existence of a complete fourth generation of light
chiral fermions in the range of 100--600~GeV to be detected at the
Large Hadron Collider (LHC) remains a very real possibility, an idea
recently revived in Ref.~\cite{Kribs:2007nz}.  As one might expect,
the masses and mixings of such new fermions must satisfy a variety of
constraints arising from direct searches, electroweak precision
constraints, and perturbative unitarity, but substantial regions in
parameter space remain for their discovery, all accessible to the
LHC\@.

In light of these phenomenological possibilities, in recent
papers~\cite{Belitsky:2010zr,Lebed:2011zg} we have developed two
4-generation models in Type IIA string theory using D6-branes
intersecting on a $T^6/(\Z_2 \times \Z_2)$ orientifold.  Our interest
in both cases was to explain the hierarchy of masses and mixings among
the known fermions.  In the first model~\cite{Belitsky:2010zr}, we
obtained full rank-4 Yukawa matrices at the level of trilinear
couplings, but fitting the large hierarchies in masses and mixings
required the adjustable vacuum expectation values (VEVs) and
open-string moduli to lie in very particular regions of parameter
space.  The second model~\cite{Lebed:2011zg}, which is of direct
interest for this paper, generates Yukawa matrices only of rank 2 at
trilinear order: Only the heaviest two generations obtain masses at
this order, while the lighter two generations obtain masses via
higher-order effects, which arguably provides a better physical
explanation for facts such as, {\it e.g.}, $m_t/m_e \approx$~340,000.
Moreover, natural solutions simultaneously accommodating the known
third-generation fermion masses and the constraints on the
fourth-generation fermion masses are relatively easy to obtain in the
model of~\cite{Lebed:2011zg}.

For the reader unfamiliar with the details of building consistent
string theory-based models containing the MSSM, the essential points
are to obtain a compactification of the extra 6 spacetime dimensions
that supports just the low-energy particle content of the theory and
to develop mechanisms for eliminating any unwanted states arising from
residual degrees of freedom.  The gauge-group charges and
multiplicities are determined by the geometry (angles and numbers of
intersections in Type IIA) through which strings wrap the compactified
dimensions.  The requirement of SUSY, the elimination of anomalies,
and the imposition of the Green-Schwarz mechanism~\cite{Green:1984sg}
to allow for massless U(1) gauge symmetries (all essential to the
topic of this paper) may all be expressed using simple well-known
constraints in terms of the wrapping numbers, brane multiplicities,
and form of the compactified space, particularly for the simple
geometry of the 6-torus orientifold $T^6/(\Z_2 \times \Z_2)$.

As mentioned above, our construction uses a Pati-Salam (PS) SU(4)$_C
\times$SU(2)$_L \times$SU(2)$_R$ scheme to obtain fermions of the
correct charges: Branes $a$, $b$, and $c$ carry color/lepton,
right-chiral, and left-chiral charges, respectively.  Unification of
the low-energy MSSM gauge group SU(3)$_C \times$SU(2)$_L
\times$U(1)$_Y$ occurs in an efficient and natural way, which in turn
is made possible because U(1)$_Y$ remains massless and anomaly free by
surviving the GS mechanism.  However, the two
models~\cite{Belitsky:2010zr,Lebed:2011zg} differ in one very
distinctive manner: In the former, the only such nonanomalous U(1)s at
the string scale (as is typical to PS constructions) are U(1)$_{B-L}$
and U(1)$_{I_{3R}}$, from which U(1)$_Y$ = $\frac 1
2$U(1)$_{B-L}$+U(1)$_{I_{3R}}$.  However, the latter model generates
an additional nonanomalous U(1)$_X$ coupling to the MSSM fields with
precisely the charges described above.  The crucial observation is
that the combinations
\begin{equation} \label{BandL}
{\rm U}(1)_B \equiv -\frac 1 4 {\rm U}(1)_X +
\frac 1 4 {\rm U}(1)_{B-L} \, , \
{\rm U}(1)_L \equiv -\frac 1 4 {\rm U}(1)_X -
\frac 3 4 {\rm U}(1)_{B-L}
\end{equation}
are nonanomalous U(1)s surviving at the string scale that, when acting
on MSSM fields, precisely count baryon and lepton number,
respectively.  These remarkable features appear due to a confluence of
a PS construction, the presence of four chiral fermion generations,
and trilinear Yukawa matrices of less-than-full rank.

In this paper we consider the broad phenomenological consequences of
the enhanced symmetries of this model.  In Sec.~\ref{sec:BL} we
exhibit beyond-MSSM operators that can contribute to proton decay,
reprise the field content of the model of Ref.~\cite{Lebed:2011zg},
and investigate which operators survive due to U(1)$_X$ conservation,
while we consider the nature of the new symmetry in
Sec.~\ref{sec:NewU1s}.  The phenomenology induced by the new symmetry
is discussed in Sec~\ref{sec:Pheno}: suppression of proton decay, the
$\mu$ term, the right-handed neutrino Majorana mass term, and
baryogenesis.  Section~\ref{sec:Prospects} emphasizes the prospects
for new physics in these models, and Sec.~\ref{sec:Concl} offers
concluding thoughts.

\section{Proton Stability and an Extra Gauged U(1)}
\label{sec:BL}

Although supersymmetry provides an elegant solution to the gauge
hierarchy problem, it introduces some additional complications.  First
among these is the rapid decay of the proton through the pair of $d =
4$ $F$-term operators ($B$- and $L$-violating, respectively):
\begin{equation}
\label{dim4operators}
U^c D^c D^c \, ,  \ Q D^c L \, .
\end{equation}
This problem is usually solved in the MSSM by introducing $R$ parity,
under which the known fermions are even while their SUSY partners are
odd (or the related ``matter parity'', under which $R = +1$ for $Q,
U^c \! , D^c \! , L, E^c \! , N^c$ and $R = -1$ for $H_{u,d}$).  As a
bonus, $R$ parity leads to a stable lightest SUSY particle (LSP),
which is a natural candidate for dark matter.  Although this idea is
attractive, it is well known that a gauged ${\rm U}(1)_{B-L}$ also
forbids the $d = 4$ operators, and furthermore, $R$ parity [more
specifically, matter parity $(-1)^{3(B-L)}$] can result from ${\rm
U}(1)_{B-L}$ broken spontaneously to its discrete $\Z_2$
subgroup~\cite{Font:1989ai, Krauss:1988zc, Martin:1992mq}.

Even though the problem of rapid proton decay via $d = 4$ operators
can be eliminated through this mechanism, one still faces the problem
of $d = 5$ operators that allow for proton decay with a lifetime too
short to evade current experimental constraints unless the
coefficients of these operators are chosen to be sufficiently small.
First among these are single operators that allow (at least in
principle) proton decay and preserve $B - L$:
\begin{equation}
\label{dim5operators}
[QQQL]_F \, , \ [U^c U^c D^c E^c]_F \, , \ [D^c D^c U^c N^c]_F \, .
\end{equation}
The second set consists of relevant $d = 5$ operators that violate
either $B$ or $L$ separately, which combine with the appropriate
member of Eq.~(\ref{dim4operators}) to form a composite operator that
conserves $B - L$ and allows proton decay:
\begin{eqnarray}
[QQQH_d]_F \, , \ [QU^cE^c H_d]_F \, , \ [QU^cL^{\dagger}]_D \, , \
[U^c (D^c)^{\dagger}E^c]_D \, , \ [QQ(D^c)^{\dagger}]_D \, ,
& & \nonumber \\
{} [QQ^{\dagger}N^c]_D \, , \ [U^c(U^c)^{\dagger}N^c]_D \, , \
[D^c(D^c)^{\dagger}N^c]_D \, , \ [QU^c N^c H_u]_F \, , \
[QD^c N^c H_d]_F \, . & &
\label{dim5operators2}
\end{eqnarray}
Indeed, the $d = 5$ operators are the ones that effectively lead to
the exclusion of GUTs based on minimal SU(5)~\cite{Murayama:2001ur},
although these operators can be suppressed in other unified models,
{\it e.g.}, flipped SU(5)~\cite{Barr:1981qv,Derendinger:1983aj}.

\begin{table}
[htb] \footnotesize
\renewcommand{\arraystretch}{1.0}
\caption{The chiral and vectorlike superfields, and their
multiplicities and quantum numbers under the gauge symmetry ${\rm
U}(4)_C\times {\rm U}(2)_L\times {\rm U}(2)_R \times \left[{\rm
USp}(8)_1 \times {\rm USp}(8)_2 \times {\rm USp}(8)_3\right]$.  The
U(1)$_X$ charge $Q_X$ is given by the combination $Q_X=
Q_4+2(Q_{2L}+Q_{2R})$.}
\label{Spectrum}
\begin{center}
\begin{tabular}{|c||c|c||c|c|c|c|c|c|c|c||c|}\hline
& Mult. & Quantum Number & $Q_4$ & $Q_{2L}$ & $Q_{2R}$ & $Q_X$ & Field
\\
\hline\hline
$ab$ & 4 & $(4,\overline{2},1,1,1,1)$ & \ 1 & $-1$ & \ 0 & $-1$ &
$F_L(Q_L, L_L)$\\
$ac$ & 4 & $(\overline{4},1,2,1,1,1)$ & $-1$ & 0 & \ 1  & \ 1 &
$F_R(Q_R, L_R)$\\
$bc$ & 4 & $(1,2,\overline{2},1,1,1)$ & 0 & 1 & $-1$  & \ 0 &
$H_u^i$, $H_d^i$\\
\hline
$a1$ & 1 & $(4,1,1,\overline{8},1,1)$ & \ 1 & 0 & 0 & \ 1 &
$X_{a1}$ \\
$a2$ & 1 & $(\overline{4},1,1,1,8,1)$ & $-1$ & 0 & 0 & $-1$  &
$X_{a2}$
\\
$b2$ & 1 & $(1,2,1,1,\overline{8},1)$ & 0 & \ 1 & 0 & \ 2 &
$X_{b2}$ \\
$b3$ & 2 & $(1,\overline{2},1,1,1,8)$ & 0 & $-1$ & 0  & $-2$  &
$X_{b3}^i$ \\
$c1$ & 1 & $(1,1,\overline{2},8,1,1)$ & 0 & 0 & $-1$  & $-2$  &
$X_{c1}$
\\
$c3$ & 2 & $(1,1,2,1,1,\overline{8})$ & 0 & 0 & \ 1  & \ 2 &
$X_{c3}^i$
\\
$b_{S}$ & 2 & $(1,3,1,1,1,1)$ & 0 & \ 2 & 0   & \ 4 &  $T_L^i$ \\
$b_{A}$ & 2 & $(1,\overline{1},1,1,1,1)$ & 0 & $-2$ & 0  & $-4$ &
$S_L^i$
\\
$c_{S}$ & 2 & $(1,1,\overline{3},1,1,1)$ & 0 & 0 & $-2$  & $-4$ &
$T_R^i$
\\
$c_{A}$ & 2 & $(1,1,1,1,1,1)$ & 0 & 0 & \ 2  & 4 & $S_R^i$ \\
\hline\hline
$ab'$ & 2 & $(4,2,1,1,1,1)$ & \ 1 & \ 1 & 0  & \ 3 & $\Omega^i_L$ \\
& 2 & $(\overline{4},\overline{2},1,1,1,1)$ & $-1$ & $-1$ & 0 & $-3$ &
$\overline{\Omega}^i_L$ \\
\hline
$ac'$ & 2 & $(4,1,2,1,1,1)$ & \ 1 & 0 & \ 1  &  \ 3 & $\Phi_i$ \\
& 2 & $(\overline{4}, 1, \overline{2},1,1,1)$ & $-1$ & 0 & $-1$ & $-3$
& $\overline{\Phi}_i$\\
\hline
$bb'$ & 4 & $(1,\overline{1},1,1,1,1)$ & 0 & $-2$ & 0  & $-4$ &
$s_L^i$ \\
      & 4 & $(1,1,1,1,1,1)$            & 0  & $ 2$ & 0 & \ 4 &
$\bar{s}_L^i$ \\
\hline
$cc'$ & 4 & $(1,1,1,1,1,1)$ & 0 & $0$ & $2$ & \ 4  & $s_R^i$ \\
      & 4 & $(1,1,\overline{1},1,1,1)$ & 0 & 0  & $-2$ & $-4$  &
$\bar{s}_R^i$ \\
\hline
$bc'$ & 1 & $(1,2,2,1,1,1)$ & 0 & \ 1 & \ 1 & \ 4 & $ H'$ \\
& 1 & $(1, \overline{2}, \overline{2},1,1,1)$ & 0 & $-1$ & $-1$ & $-4$
& $\overline{H}'$\\
\hline
\end{tabular}
\end{center}
\end{table}

Let us consider the model constructed in Type IIA string theory with
intersecting D6-branes discussed in~\cite{Lebed:2011zg}. At the string
scale, it is a four-generation Pati-Salam (PS) model, with the gauge
symmetry of the ``observable'' sector given by
\begin{equation}
{\rm U}(4)_C \times {\rm U}(2)_{L} \times {\rm U}(2)_{R} \, , 
\end{equation}
in addition to a ``hidden'' sector with the gauge group ${\rm
USp}(8)^3$.  Since U($N$) = SU($N$)$\times$U(1), each observable gauge
group has an associated anomalous U(1), denoted here as ${\rm
U}(1)_4$, ${\rm U}(1)_{2L}$, and ${\rm U}(1)_{2R}$, respectively
(which correspond to the $a$, $b$, and $c$ brane stacks,
respectively).  The anomalies of these Abelian symmetries are
cancelled by a generalized Green-Schwarz mechanism, and as a result
their gauge bosons obtain string-scale masses.  However, these U(1)s
remain as global symmetries to all orders in perturbation theory, and
as a result perturbatively forbid operators that would otherwise be
allowed.

As noted in~\cite{Lebed:2011zg}, precisely one linear combination of
${\rm U}(1)_4$, ${\rm U}(1)_{2L}$, and ${\rm U}(1)_{2R}$ remains
massless and anomaly free, which is given by
\begin{equation}
{\rm U}(1)_X \equiv {\rm U}(1)_{4} + 2\left[{\rm U}(1)_{2L} +
{\rm U}(1)_{2R} \right] \, .
\end{equation}
Thus, the full gauge symmetry of the model at this stage is given by
\begin{equation}
{\rm SU}(4)_C \times {\rm SU}(2)_L \times {\rm SU}(2)_R \times {\rm
U}(1)_X \times \left[{\rm USp}(8)^3\right].
\end{equation}
We exhibit the matter spectrum and the quantum numbers of each field
in Table~\ref{Spectrum}.  As seen there, the superfields $F_L^i(Q_L,
L_L)$ carry charge U(1)$_X$ charge $Q_X = -1$, $F_R^i(Q_R, L_R)$
(= $U^c$, $D^c$, $N^c$, $E^c$) carry charge $Q_X = +1$, and the Higgs
superfields are uncharged under U(1)$_X$, as promised in the
Introduction.  Thus, the Yukawa couplings
\begin{eqnarray}
W_Y & = & y_{ijk} F^i_L F^j_R H^k \\ \nonumber 
    & = & y^u_{ijk} Q^i (U^c)^j H_u^k +
          y^d_{ijk} Q^i (D^c)^j H_d^k +
          y^{\nu}_{ijk} L^i (N^c)^j H_u^k +
          y^l_{ijk} L^i (E^c)^j H_d^k \, ,
\end{eqnarray}
where $i,j,k = \left\{1,2,3,4\right\}$, are allowed by both the global
U(1) symmetries as well as the gauged ${\rm U}(1)_X$ symmetry.  As
shown in~\cite{Lebed:2011zg}, the resulting Yukawa matrices are rank
2, which allows for fermion mass textures that can easily satisfy
constraints placed on fourth-generation fermion masses.  Since the
fields $F_L$ and $F_R$ have different U(1)$_X$ charges, these fields
obviously cannot be placed in the same spinorial $\mathbf{16}$ of
SO(10).  Thus, this PS model cannot have an SO(10) origin.

As can also easily be checked using Table~\ref{Spectrum}, none of the
$d = 4$ and $d = 5$ operators given in Eqs.~(\ref{dim4operators}),
(\ref{dim5operators}), and (\ref{dim5operators2}) are singlets under
${\rm U}(1)_X$ (Nor, it should be added, are any of the remaining 21
operators of the set of 36 mentioned in the Introduction, all of which
are purely $L$-violating).  Indeed, since U(1)$_X$($Q$) = +1,
U(1)$_X$($U^c$, $D^c$) = $-1$, U(1)$_X (H_u, H_d) = 0$, one finds that
it is not possible to create a $B$-violating U(1)$_X$-singlet $d
\le 6$ operator without including exotic matter.  Thus, $B$- and
$L$-violating processes that proceed via these operators are
effectively forbidden at scales where U(1)$_X$ remains unbroken.  

As mentioned in the previous section, the MSSM superfields are also
charged under global symmetries that arise from anomalous U(1)s whose
gauge bosons obtain masses via the generalized Green-Schwarz anomaly
cancellation mechanism.  As can be seen from Table~\ref{Spectrum},
none of the dimension-4 and -5 operators that mediate proton decay are
singlets under these global symmetries, and are thus perturbatively
forbidden.  However, these global symmetries can, in principle, be
broken by nonperturbative effects, namely D-brane
instantons~\cite{Cvetic:2006iz,Cvetic:2009ng,Blumenhagen:2009qh},
assuming that suitable instantons with the correct zero-mode structure
and satisfying all constraints can actually arise.
  
If such suitable instantons are present in the model, then proton
decay can be allowed via these operators, albeit at a rate
exponentially suppressed by the instanton action.  While it might be
possible for the instanton suppression to be large enough to allow
proton decay at an acceptable rate, this scenario is by no means
guaranteed.  However, for the present model, these operators cannot be
induced via D-brane instantons since the U(1)$_X$ gauge symmetry
remains unbroken at the string scale.  Therefore, proton decay via
these operators is forbidden at scales where U(1)$_X$ remains
unbroken, and highly suppressed at scales at which U(1)$_X$ is broken
since the effects by which they may appear are perturbatively
forbidden.

\section{The U(1)$_L$- and U(1)$_B$-Extended MSSM's}
\label{sec:NewU1s}

The PS gauge symmetry may be broken by a process of D6-brane
splitting~\cite{Cvetic:2004nk,Cvetic:2004ui}, which corresponds to
assigning VEVs to some of the adjoint scalars associated with each
stack along its flat directions (See~\cite{Lebed:2011zg} for a more
detailed discussion of the process for this model).  For present
purposes, we note that the splitting results in additional massless
and anomaly-free U(1)s given by
\begin{eqnarray}
{\rm U}(1)_{B-L} & = & \frac{1}{3}{\rm U}(1)_{\rm{baryon}} -
{\rm U}(1)_{\rm{lepton}} \, ,
\nonumber \\
{\rm U}(1)_{I_{3R}}& = & \frac{1}{2}{\rm U}(1)_{\rm{up}} -
\frac{1}{2}{\rm U}(1)_{\rm{down}} \, , 
\label{BLI3R}
\end{eqnarray}
while ${\rm U}(1)_X$ becomes 
\begin{eqnarray}
{\rm U}(1)_X & = & {\rm U}(1)_{\rm{baryon}} +
{\rm U}(1)_{\rm{lepton}} +
2\left[ {\rm U}(1)_{2L} + {\rm U}(1)_{\rm{up}} +
{\rm U}(1)_{\rm{down}} \right] \nonumber \\
& = & {\rm U}(1)_{a1} +
{\rm U}(1)_{a2} +
2\left[ {\rm U}(1)_{b} + {\rm U}(1)_{c1} +
{\rm U}(1)_{c2} \right],
\label{X}
\end{eqnarray}
where ${\rm U}(1)_{\rm baryon}$=${\rm U}(1)_{a1}$ and ${\rm
U}(1)_{\rm{lepton}}$=${\rm U}(1)_{a2}$ are global U(1)s obtained from
U(1)$_4=$U(1)$_a$ after brane splitting, under which baryons and
leptons are charged, respectively.  Similarly, ${\rm
U}(1)_{\rm{up}}$=${\rm U}(1)_{c1}$ is a global symmetry under which
up-type quarks and neutrinos are charged, while down-type quarks and
electrically-charged leptons are charged under the global symmetry
$\rm{U}(1)_{\rm{down}}$=${\rm U}(1)_{c2}$, both obtained from
U(1)$_{2R}=$U(1)$_b$ after brane splitting.  Typically in this type of
model, the charges of the quarks and leptons under these global (and
indeed, anomalous) symmetries are identified as $\frac 1
3$U(1)$_{\rm{baryon}}$ and U(1)$_{\rm{lepton}}$, respectively.
However, as we now show, in this model one finds it is possible to
obtain {\it gauged\/} U(1)s under which the quarks and leptons have
numerically identical charges, that we label as the true baryon number
($B$) and lepton number ($L$), respectively.  In particular, when
acting upon MSSM fields, the U(1)$_X$ gauge symmetry is effectively
$-{\rm U(1)}_{3B+L}$.

After brane splitting, the observable gauge symmetry of the model
becomes ${\rm SU}(3)_C \times {\rm SU}(2)_L \times {\rm U}(1)_{B-L}
\times {\rm U}(1)_{I_{3R}} \times {\rm U}(1)_{3B+L}$, with
${\rm U}(1)_{B-L}$, ${\rm U}(1)_{I_{3R}}$, and ${\rm
U}(1)_{3B+L}=-{\rm U(1)}_X$ as defined in Eq.~(\ref{BLI3R}).  Since
both ${\rm U(1)}_{B-L}$ and ${\rm U(1)}_{3B+L}$ remain good
symmetries, one immediately sees why none of the $B$ and $L$-violating
operators that mediate proton decay in the MSSM are allowed:
Equation~(\ref{BandL}) is equivalent to the trivial identities
\begin{eqnarray}
{\rm U}(1)_B = \frac{1}{4}[{\rm U}(1)_{B-L} + {\rm U}(1)_{3B+L}] \, ,
\ \ \ \ \ \
{\rm U}(1)_L = \frac{1}{4}[-3{\rm U}(1)_{B-L}+ {\rm U}(1)_{3B+L}] \, , 
\label{GaugedBL}
\end{eqnarray}
that, when acting on MSSM fields, serve the promised role of precisely
counting baryon and lepton number, respectively.  The relation between
the global and gauged forms is given by
\begin{eqnarray}
{\rm U}(1)_B & = & {\rm U}(1)_{\rm baryon} + {\rm U}(1)_{\Delta} \, ,
\nonumber \\
{\rm U}(1)_L & = & {\rm U}(1)_{\rm lepton} + {\rm U}(1)_{\Delta} \, ,
\end{eqnarray}
where
\begin{equation}
{\rm U}(1)_{\Delta} = -\frac 1 2 \left[ 3{\rm U}(1)_{\rm baryon} +
{\rm U}(1)_{\rm lepton} + {\rm U}(1)_{2L}
+ {\rm U}(1)_{\rm up} + {\rm U}(1)_{\rm down} \right]
\end{equation}
is also an anomalous U(1), but under which no MSSM field is charged. 

The fact that two equivalent combinations of U(1)$_{B-L}$ and
U(1)$_{3B+L}$ precisely count $B$ and $L$ indicates that both U(1)$_B$
and U(1)$_L$ remain gauged as long as both parent symmetries remain
unbroken.  One may ask how the full gauge symmetry may be further
broken to yield the individual combinations shown in
Eq.~(\ref{GaugedBL}).  As an example, first note that the ${\rm
U(1)}_{B-L} \times {\rm U(1)}_{I_{3R}}$ gauge symmetry may be broken
to the SM hypercharge ${\rm U(1)}_Y$ by assigning VEVs to the
vectorlike singlet fields $\Phi$, $\overline{\Phi}$ (from the $a_2
c_2'$ intersections) carrying the quantum numbers $({\bf 1}, {\bf 1},
\frac 1 2, -1, -3)$ and $({\bf 1}, {\bf 1}, -\frac 1 2, 1, 3)$ under
the ${\rm SU}(3)_C\times {\rm SU}(2)_L\times {\rm U}(1)_{I_{3R}}
\times {\rm U}(1)_{B-L} \times {\rm U}(1)_{3B+L}$ gauge symmetry.  Since
$\Phi$, $\overline{\Phi}$ carry nonzero charges under both
U(1)$_{B-L}$ and U(1)$_{I_{3R}}$, the two gauge symmetries are broken,
but the fields are uncharged under the linear combination
\begin{eqnarray}
{\rm U}(1)_Y & = & \frac{1}{6}\left[{\rm U}(1)_{\rm baryon}
- 3{\rm U}(1)_{\rm lepton}+ 3{\rm U}(1)_{\rm up}
- 3{\rm U}(1)_{\rm down}\right] \\ \nonumber
& = & \frac{1}{6}\left[{\rm U}(1)_{a1}- 3 {\rm U}(1)_{a2}
+ 3 {\rm U}(1)_{c1} -3 {\rm U}(1)_{c2}\right] \\ \nonumber
& = & \frac{1}{2} {\rm U}(1)_{B-L} + {\rm U}(1)_{I_{3R}} \, ,
\end{eqnarray}
corresponding to the SM hypercharge, which then remains unbroken.  In
addition, these vectorlike fields carry charges under the
U(1)$_{3B+L}$ gauge symmetry and so it is also broken.  However, these
fields do not carry charges under the linear combination corresponding
to U(1)$_L$ given in Eq.~(\ref{GaugedBL}), and thus it survives.
Therefore, at this stage, the gauge symmetry of the observable sector
is given by
\begin{equation}
{\rm SU}(3)_C \times {\rm SU}(2)_L \times {\rm U}(1)_Y \times
{\rm U}(1)_L \, , 
\label{MSSM_L}
\end{equation}
just the gauge symmetry of the MSSM, extended by an extra U(1) that
counts $L$.

Many alternate scenarios for symmetry breaking are possible.  For
example, the U(1)$_{B-L}\times$U(1)$_{I_{3R}}\times$U(1)$_{3B+L}$
gauge symmetry may instead be broken by assigning VEVs to the
right-handed neutrino fields $N_R$.  In this case, the gauge symmetry
is broken to
\begin{equation}
{\rm SU}(3)_C \times {\rm SU}(2)_L \times {\rm U}(1)_Y \times
{\rm U}(1)_B \, . 
\label{MSSM_B}
\end{equation}
However, assigning VEVs to $N_R$ breaks SUSY, which is expected not to
occur until the TeV scale.  To avoid such outcomes, the
U(1)$_{I_{3R}}\times$U(1)$_{B-L} \times$U(1)$_{3B+L}$ symmetry
breaking is thus naively expected to occur near the usual GUT scale.
Indeed, as shown in~\cite{Lebed:2011zg}, if the PS gauge symmetry is
broken to the MSSM at the GUT/string scale, then the MSSM gauge
couplings are unified.  Thus, it is natural to assume that the $\Phi$,
$\overline{\Phi}$ fields from the $a_2c'_2$ sector obtain GUT-scale
VEVs.  That being said, the possibility of breaking the gauge symmetry
at the TeV-scale by assigning VEVs to $N_R$ and not $\Phi$,
$\overline{\Phi}$ is worth exploring.

\section{Phenomenological Constraints on Singlet VEVs}
\label{sec:Pheno}

In the previous section we demonstrated that the gauge symmetry may be
broken to ${\rm SU}(3)_C \times {\rm SU}(2)_L \times {\rm U}(1)_Y
\times {\rm U}(1)_L$ at the GUT scale by assigning VEVs to the
vectorlike fields $\Phi$, $\overline{\Phi}$, or to ${\rm SU}(3)_C
\times {\rm SU}(2)_L \times {\rm U}(1)_Y \times {\rm U}(1)_B$ at the
TeV scale by assigning VEVs to the right-handed neutrinos $N_R$.
These two cases show that models of this type may be adapted to
provide either a U(1)$_L$ or U(1)$_B$ that survives unbroken to low
energies.

However, other singlet fields appear in the model whose VEVs may break
U(1)$_{3B+L}$ [or equivalently, U(1)$_B$ and U(1)$_L$] at intermediate
scales, namely, the singlets $S_L$ and $S_R$, as well as the SU(2)$_R$
triplet fields $T_R$.  In order to determine the scales at which these
fields may obtain VEVs, one must study other observable phenomena
mediated by these fields.  As we shall see below, one
can construct operators that allow proton decay, which imposes
constraints upon the VEVs of the $S_R$ fields.  In addition, one can
construct superpotential operators for the Higgsino bilinear $\mu$
term and for the Majorana mass term for right-handed neutrinos, which
are in turn constrained.

\subsection{\bf Proton Decay via Exotic Fields}

Although the $d = 4$ and $d = 5$ operators involving only MSSM fields
that could mediate proton decay are excluded by the gauged
U(1)$_{B-L}$ and U(1)$_{3B+L}$ gauge symmetries, the model also
contains fields that are not part of the MSSM, yet are charged under
U(1)$_B$ and/or U(1)$_L$.  Operators involving these extra fields can
mediate $B$- or $L$-violation.  Furthermore, $B$- and $L$-violation
can occur below energy scales at which U(1)$_{B-L}$ and U(1)$_{3B+L}$
are spontaneously broken [and, as discussed previously, if such
operators are perturbatively forbidden by the global U(1) symmetries,
then additional instanton suppression factors appear].  One must
examine the possible operators that arise and their VEVs in order to
fully address these issues.

The leading such operator involving exotic fields that mediates proton
decay is
\begin{equation}
\frac{\kappa}{M_{\rm St}^2}~QQQLS_R \, ,
\end{equation}
where $M_{\rm St} = O(10^{18}~{\rm GeV})$ is the string scale.  Even
though it is a singlet under all gauge groups, this operator is
perturbatively forbidden by the global symmetries U(1)$_{4,2L,2R}$ and
can only be induced via nonperturbative effects such as from D-brane
instantons, which introduce the suppression factor $\kappa
\propto e^{-S_{E2}}$.  This operator is effectively $d = 5$, assuming
that the singlet fields $S^i_R$ obtain VEVs:
\begin{equation}
\label{leadingpdecayop}
\frac{e^{-S_{E2}} \left\langle S^i_R\right\rangle}
{M_{\rm St}^2}~QQQL \, .
\end{equation}
Using the phenomenological bound ({\it e.g.}, \cite{Ibanez:1991pr}) on
the operator $QQQL$, we require
\begin{equation}
\label{pdecayconstraint}
\frac{e^{-S_{E2}} \left\langle S^i_R\right\rangle}{M_{\rm St}} \leq
O(10^{-7}) \, ,
\end{equation}
in order to satisfy the experimental lower limit on the proton
lifetime.  This estimate, of course, assumes the existence of a
suitable D-brane instanton with the appropriate zero-mode structure in
the model, which is far from obvious.  Assuming such an instanton
exists in the worst-case scenario $e^{-S_{E2}}\approx 1$, then the
maximum allowed VEV is $\left\langle S_R \right\rangle \approx
10^{11}$~GeV; at the other extreme, an instanton suppression of
$10^{-7}$ would allow $\langle S_R^i \rangle$ to be as large as
$M_{\rm St}$.  A thorough analysis, which we do not undertake here,
would require enumerating operators that, upon receiving VEVs,
effectively assume the forms of the ``dangerous'' operators in
Eqs.~(\ref{dim4operators})--(\ref{dim5operators2}); however, the
approach is completely analogous to that for
Eq.~(\ref{leadingpdecayop}).

Of course, the most straightforward way to eliminate the possibility
of rapid proton decay via such operators involving exotic fields such
is to forbid these fields from obtaining large VEVs.  However, other
phenomenological properties can depend on VEVs of these fields.  Among
these, in particular, are the $\mu$ term and the Majorana mass term
for right-handed neutrinos, which we discuss next.

\subsection{The $\mu$ Term and Majorana Neutrino Masses}

The problem of generating a $\mu$ superpotential term of the form $\mu
H_u H_d$, with $\mu= O$(TeV), is well known in supersymmetric models.
Specifically, the question of why the $\mu$ parameter is TeV-scale
rather than Planck-scale has long been an open question in the MSSM\@.
In the context of intersecting D-brane models, this problem is
somewhat ameliorated in that a simple bilinear coupling is forbidden
by global U(1) symmetries [here, U(1)$_{2L,2R}$].  Such a term can
therefore only be generated either by high-order couplings or by
nonperturbative effects such as D-brane
instantons~\cite{Blumenhagen:2009qh}.  In either case, one expects the
effective operator to be suppressed by powers of $M_{\rm St}$.

One possibility for generating a $\mu$ term is through the
higher-dimensional ($d = 5$) operator
\begin{eqnarray}
\label{mu-term}
W & \supset & \frac{y^{ijkl}_{\mu}}{M_{\rm St}} S_L^i S_R^j
H_u^k H_d^l  ~,~\,
\end{eqnarray}
where $y^{ijkl}_{\mu}$ are Yukawa couplings.  In this case, the
singlets $S_R$ may obtain string or GUT-scale VEVs (or lower, as
described previously) while preserving the D-flatness of U(1)$_{2R}$,
and the singlets $S_L$ may obtain TeV-scale VEVs while preserving the
D-flatness of U(1)$_{2L}$, while the Higgses couple through their
electroweak-scale VEVs.  Simple order-of-magnitude estimates then show
that a TeV-scale $\mu$ term may be generated, with $y^{ijkl}_{\mu} =
O(1)$.

For a $\mu$ term of the desired magnitude to be generated by this
higher-dimensional operator, the fields $S_R$ must obtain VEVs near
$M_{\rm St}$.  However, as discussed in the previous section, the
constraint $\langle S_R^i \rangle \! \ll \! M_{\rm St}$ is expected
from proton decay unless strong instanton suppressions are present.
Thus, while this class of models does provide a mechanism to generate
$\mu$ of natural size, it is somewhat preferable for $S_R$ to obtain
VEVs at a scale much lower than $M_{\rm St}$, and for the (bulk of
the) $\mu$ term to be generated via some other mechanism, such as
D-brane instantons.

From Table~I and Eqs.~(\ref{BLI3R}), (\ref{X}), and (\ref{GaugedBL}),
one may check that $S_R^i$ is a singlet under $B - L$ but is charged
under $3B + L$, and therefore $\langle S_R^i \rangle$ spontaneously
breaks U(1)$_L$ (and similarly for $\langle T_R^i \rangle$).  One then
expects the scale at which $S_R$ and $T_R$ obtain VEVs to be
correlated with the scale at which the right-handed neutrinos obtain
Majorana masses, an idea that is implemented as follows.  A simple
Majorana mass term for right-handed neutrinos of the form $M \!
\cdot \!  N^c N^c$ is perturbatively forbidden in the model by
the global U(1) symmetries [here, U(1)$_{4,2R}$].  A Majorana mass can
therefore only be generated by high-dimensional operators or by
nonperturbative effects.  Indeed, the possibility of generating a
Majorana mass term via D-brane instantons has been much studied in the
literature~\cite{IbaUra06,BluCveWei06}.  However, it should be
mentioned that for the models studied in~\cite{Belitsky:2010zr,
Lebed:2011zg, Chen:2007px, Chen:2007zu}, a standard Majorana mass term
$M \! \cdot \! N^c N^c$ is not a singlet under U(1)$_{B-L}$ [or, for
that matter, U(1)$_{3B+L}$] and therefore cannot be generated solely
via D-brane instantons.  On the other hand, Majorana masses can be
generated in these models without instantons by high-dimensional
operators such as
\begin{eqnarray}
W & \supset & \frac{y^{mnkl}_{Nij}}{M^3_{\rm St}} T_R^{m}
T_R^{n} \Phi^i \Phi^j  F_R^k  F_R^l ~,~\,
\label{eqn:HiggsSup}
\end{eqnarray}
where $y^{mnkl}_{Nij}$ are Yukawa couplings, and the fields $T_R$ and
$\Phi$ may obtain VEVs at the string scale (or lower).  For the model
of Table~\ref{Spectrum}, right-handed neutrino masses can be generated
in the range $10^{10-14}$~GeV for $y^{mnkl}_{Nij} \sim
10^{(-7)-(-3)}$, assuming GUT- or string-scale VEVs for the $\Phi$ and
$T_R$.  As noted above, $\langle T_R^i \rangle$ also break ${\rm
U}(1)_{3B+L}$ and U(1)$_L$, which suggests that lower values of
$\langle T_R^i \rangle$ (comparable to or perhaps somewhat larger than
those of $\langle S_R^i \rangle$, say, $10^{14}$~GeV) allow for
$y^{mnkl}_{Nij}$ to be of a natural $O(1)$ size.

In summary, if the $\mu$ term and the Majorana mass term are generated
by high-dimensional operators involving the fields $S_R$ and $T_R$
with string-scale VEVs, then U(1)$_{3B+L}$ is broken near $M_{\rm
St}$.  In this case, the only Abelian factor that survives at low
energies is the weak hypercharge U(1)$_Y$, and the suppression of
proton decay and the Majorana mass scale require small coefficients,
but the $\mu$ term is of natural size.  If $\langle S_R^i \rangle$ and
$\langle T_R^i \rangle$ lie closer to the Majorana mass scale, then
proton decay and Majorana mass terms can have natural coefficients,
but then the $\mu$ term must be generated by effects other than
high-dimensional operators, in particular via D-brane
instanton-induced operators.  Thus, it is quite possible that the
$\mu$ term and Majorana mass term may be generated without requiring
string-scale VEVs for the fields $S_R$ and $T_R$.  If this is the
case, then U(1)$_L$ can remain unbroken down to the electroweak scale.
 
\subsection{Baryogenesis}
The existence of $B$-violating processes is one of the three Sakharov
conditions~\cite{Sakharov:1967dj} that must be satisfied in order for
baryogenesis to be realized.  As shown for this model, processes that
can mediate proton decay are strongly suppressed.  Indeed, the
relevant operators are forbidden by U(1)$_X$ and/or perturbatively
forbidden by global U(1) symmetries.  In fact, if global U(1)$_{\rm
baryon}$ must be conserved [{\it i.e.}, if no nonperturbative
U(1)$_{\rm baryon}$-violating interaction can be generated] and the
nonzero VEVs are the aforementioned $\left< \Phi_{a2c2'} \right>$,
$\left< S_L \right>$, $\left< S_R \right>$, and $\left< T_R \right>$
[all of which are singlets under U(1)$_{\rm baryon}$], $B$-violation
is forbidden.  This result holds for both the single-operator and
two-operator proton-decay combinations.  The proton is effectively
stable under these assumptions.  On the other hand, since all of the
VEVs listed above break global U(1)$_{\rm lepton}$, one can show that
{\it every\/} $L$-violating, $B$-conserving operator out of the 36 can
be produced with the four given VEVs without violating any symmetry.
Thus, $L$-violating processes clearly dominate over $B$-violating ones
in this model.

The most likely possibility for realizing the matter-antimatter
symmetry of the universe within this model is therefore electroweak
baryogenesis~\cite{Fukugita:1986hr,Trodden:1998ym}.  In this scenario,
leptogenesis occurs first, thereby generating a lepton-number
asymmetry.  Nonperturbative field configurations known as sphalerons
then convert this lepton-number asymmetry into a baryon-number
asymmetry, resulting in baryogenesis.  Our model allows numerous
$L$-violating operators at the perturbative level, while $B$-violation
requires instanton-like configurations to occur; even if such
configurations are heavily suppressed in the modern universe, they may
be much easier to realize in the high-temperature environment of the
early universe.  Electroweak baryogenesis is therefore touted as a
natural means of generating a baryon asymmetry while also maintaining
the effective stability of the proton.

One should recall that a sufficient amount of CP violation (CPV) is
also a necessary Sakharov condition for baryogenesis.  While the
minimal three-generation SM is well known to possess insufficient CPV
to account for baryogenesis, the introduction of a fourth
generation~\cite{Hou:2010mm} provides a suitable natural source of
CPV\@.  Other CPV sources enter through the MSSM two-Higgs potential,
and indeed, through the couplings of numerous exotic fields appearing
in our string-inspired model.  All of the basic ingredients necessary
to realize the observed baryogenesis therefore appear to be present in
this model.

\section{Prospects for New Physics}
\label{sec:Prospects}

The four-generation model (or more accurately, family of models)
discussed here was constructed in Type IIA string theory with
D6-branes intersecting at angles.  As such, the model satisfies all
global consistency constraints as well as preserving $\mathcal{N}=1$
supersymmetry.  In addition, the model has a rich phenomenology, which
includes rank-2 Yukawa matrices that can provide a natural explanation
for the mass textures required to satisfy current constraints on a
fourth generation of chiral fermions.  Also, the MSSM gauge couplings
are unified at the tree-level in the model, and matter charged under
the hidden sector decouples at high energies.  These features by
themselves make this string theory model of high phenomenological
interest, and provide additional motivation for the possible existence
of a fourth generation of quarks and leptons that might be observed at
the LHC.

Given the above features, it is really quite remarkable that the model
automatically possesses an extra nonanomalous U(1) gauge symmetry that
forbids all of the dimension-4, -5, and -6 operators that mediate
proton decay in the MSSM\@.  Although the idea of eliminating these
operators using an extra nonanomalous U(1) is not new, the extra U(1)
of this model is not added to the model by hand, but rather emerges
automatically from the D-brane construction.  Specifically, this extra
U(1) has not been made anomaly-free by arbitrarily adding extra matter
representations into the model as is done in most phenomenological
models in the literature.  Among the phenomenological consequences of
the existence of this U(1) is that the proton is, for all intents and
purposes, stable in the model.  This result is consistent with the
non-observation of proton decay, and furthermore implies that proton
decay should not be observed in future experiments.  This being said,
as discussed in the previous section, proton decay may be allowed in
principle via operators involving exotic matter fields.  However,
these operators are forbidden by global symmetries in the model, and
it has not been demonstrated that they may be induced by
nonperturbative effects such as D-brane instantons.  Even if such
effects occur, one then merely obtains constraints on the VEVs of the
exotic singlet fields in the model, as discussed above.

As the dimension-4 operators that mediate rapid proton decay are
forbidden by the Abelian gauge symmetries, the role for an $R$ parity
is much diminished in the model; $R$ parity is not required to
eliminate rapid proton decay in this model as it does in the MSSM\@.
This fact may have dramatic consequences for the supersymmetric
phenomenology at the LHC and for dark matter experiments.  In
particular, the absence of $R$ parity results in an unstable Lightest
Supersymmetric Particle (LSP).  In the MSSM with $R$ parity, the LSP
is stable and thus must be electromagnetically neutral, which provides
a natural candidate for dark matter.  In addition, many of the
collider signatures that provide the cleanest signals for the
production of supersymmetric particles are those with large missing
transverse energy (MET).  Such signals are expected to arise when the
produced supersymmetric particles decay directly to the LSP, which is
typically a neutralino.  However, if the LSP is no longer stable in
the absence of $R$ parity, then these MET signals do not have the same
significance.  Indeed, in this scenario the LSP is not even required
to be electromagnetically neutral.  The lack of $R$ parity makes the
observation of supersymmetric particles at the LHC a much more
difficult proposition.  In addition, the LSP cannot be stable dark
matter, which would strongly impact the prospects for direct dark
matter detection experiments.

Finally, as we have seen, it is possible for the gauge symmetry to be
broken to the MSSM with an extra gauged U(1) that couples either to
lepton [U(1)$_L$] or to baryon [U(1)$_B$] number.  Whether one obtains
a gauged lepton number or a gauged baryon number at low energies
depends intrinsically upon which fields are used to break the gauge
symmetry.  Either U(1)$_L$ or U(1)$_B$ may survive unbroken to the TeV
scale or below.  If so, the symmetry would be heralded by so-called
leptophilic or leptophobic $Z'$ bosons, respectively, which would be
observable at the LHC\@.  As mentioned above, a leptophilic $Z'$ could
explain~\cite{Fox:2008kb} the observed
PAMELA~\cite{Adriani:2008zr}/ATIC~\cite{Chang:2008zzr}/Fermi
LAT~\cite{Ackermann:2011rq} cosmic ray positron excess, while a
leptophobic $Z'$ has been discussed in connection with recent
anomalies observed by the CDF
collaboration~\cite{Buckley:2011vc,Aaltonen:2011kc,Aaltonen:2011mk}.
Such considerations deserve detailed study, which we leave for future
work.

\section{Conclusions} \label{sec:Concl}

We have demonstrated the existence of an extra massless and
anomaly-free U(1) gauge symmetry in the model constructed
in~\cite{Lebed:2011zg} that plays a central role in suppressing $B$-
and $L$-violation.  In particular, we obtained linear combinations of
this U(1)$_X$ with ${\rm U}(1)_{B-L}$ that count baryon and lepton
number, respectively, thereby providing U(1)$_{B,L}$ gauge symmetries
at the string scale that forbid all dimension-4, dimension-5, and
dimension-6 operators involving only MSSM fields that mediate proton
decay.

Nevertheless, additional fields charged under the gauged $B$ and $L$
symmetries appear in the model; other operators involving this exotic
matter arise that can mediate proton decay through their VEVs, once
U(1)$_X$ breaks spontaneously.  We found the leading such operator
$QQQLS_R$, and while noting that it can only appear nonperturbatively
in the model, estimated the constraint on the operator's coefficient.
Constrained, in turn, is the scale at which the singlet fields $S_R$
may receive VEVs.  We noted that VEVs for $S_R$ break U(1)$_L$; if
$\left< S_R \right>$ is low, as suggested by the proton decay
constraint, then U(1)$_L$ can potentially remain unbroken down to the
electroweak scale.  If so, this scenario can provide a so-called
leptophilic $Z'$ boson, which could be observed at the LHC and which
can explain the galactic cosmic ray positron excess observed by the
PAMELA collaboration.

The appearance of this extra massless and anomaly-free gauge symmetry
U(1)$_X$ in the model is quite remarkable.  This extra U(1) occurs
naturally alongside the usual MSSM gauge symmetry, and is therefore
arguably a natural extension of the MSSM\@.  The fact that U(1)$_X$
forbids all of the well-known dimension-4 and -5 operators that can
mediate rapid proton decay, while also allowing all of the Yukawa
couplings for quarks and leptons, is especially remarkable.  As we
have seen, this extra Abelian gauge symmetry can be viewed as a means
of gauging baryon and lepton number, which is strongly compatible with
the non-observation of $B$- and $L$-violating effects such as proton
decay and neutrinoless double beta decay.  These features, combined
with previous results shown for this particular model that the MSSM
gauge couplings unify at the string scale and that it is possible to
naturally obtain realistic Yukawa mass matrices for quarks and
leptons, give this model particular phenomenological value.

\section*{Acknowledgments}
This work was supported in part by the National Science Foundation
under Grant No.\ PHY-0757394.  RFL also thanks the Galileo Galilei
Institute for Theoretical Physics for their hospitality and the INFN
for partial support.

\end{document}